\documentclass[a4paper,11pt]{article}
\usepackage{pos}
\usepackage{booktabs}
\usepackage{multicol}
\usepackage{topcapt}
\usepackage{xspace}

\title{The High-Level Trigger for the CMS Phase-2 Upgrade}
\ShortTitle{The High-Level Trigger for the CMS Phase-2 Upgrade}

\manuallySeparateAuthors
\author*{Thiago Rafael Fernandez Perez Tomei}
\author{ on behalf of the CMS Collaboration}

\affiliation{Universidade Estadual Paulista,\\
  São Paulo, SP, Brazil}

\emailAdd{Thiago.Tomei@cern.ch}

\abstract{The High-Luminosity LHC (HL-LHC) will usher {in} a new era in high-energy physics. The HL-LHC experimental conditions entail an instantaneous luminosity of up to {$7.5 \times 10^{34}\,\mathrm{cm}^{-2}\,\mathrm{s}^{-1}$} and up to 200 simultaneous collisions per bunch crossing (pileup). To cope with those conditions, the CMS detector will undergo a series of improvements, in what is known as the Phase-2 upgrade. In particular, the upgrade of the Data Acquisition and of the High-Level Trigger (DAQ--HLT) will have to address a much higher event rate and more complex events. In this paper, we will discuss the aspects of the HLT upgrade, detailing the development of the online reconstruction, the construction, characterisation and timing/rate measurement of a simplified HLT menu, the role of heterogeneous architectures in the HLT and the plan of work and milestones until the beginning of Phase-2.}

\FullConference{%
  41st International Conference on High Energy physics - ICHEP2022\\
  6-13 July, 2022\\
  Bologna, Italy
}

\newcommand{\PT}{\ensuremath{p_{\mathrm{T}}}\xspace}
\newcommand{\HT}{\ensuremath{H_{\mathrm{T}}}\xspace}
\newcommand{\pileupof}[1]{$\langle{\rm PU}\rangle = #1$\xspace}

\renewcommand{\logo}{\relax}
\renewcommand{\hookAfterAbstract}{\par\bigskip \textsc{ArXiv ePrint}: \href{https://arxiv.org/abs/2211.03684}{2211.03684} }

\begin{document}
\maketitle

%%%%%%%%%%%%%%%%
\section{Introduction}

The Compact Muon Solenoid (CMS) detector~\cite{CMS:2008xjf} will undergo the so-called Phase-2 upgrade~\cite{CMSCollaboration:2015zni}
{and this will} allow the experiment to address the conditions of the High-Luminosity LHC (HL-LHC).
All CMS subdetectors will be upgraded, and the Trigger and Data Acquisition system is no exception.
The experiment will continue using a two-level trigger system, similarly to what {was} done for the first LHC era.
The Phase-2 Level-1 Trigger (L1T) will be implemented in custom-made electronics and dedicated to analyse the detector information 
%at a coarse-grained scale, while
with a maximum latency of 12.5\,\textmu{}s;
the Phase-2 High Level Trigger (HLT) will be implemented as a series of software algorithms, running in a heterogeneous-architecture computing farm.
%, that have access to the full detector information.
The complete information on the Phase-2 HLT is available in the DAQ--HLT Technical Design Report (TDR)~\cite{Collaboration:2759072}.

The Phase-2 HLT will analyse the full L1T output of up to 750\,kHz. 
It will need high selection efficiency for events of interest 
whilst staying within the envelopes of acceptable output rate / bandwidth and computing farm size dictated by the experimental budget.
Table~\ref{tab:1} summarises the CMS Phase-2 trigger and DAQ projected running parameters.

\begin{table}[htbp]
\centering
\scalebox{0.72}{\begin{tabular}{lccc}
\toprule 
& LHC & \multicolumn{2}{c}{ HL-LHC } \\ 
CMS detector & Phase-1 & \multicolumn{2}{c}{ Phase-2 } \\ 
Peak \(\langle\mathrm{PU}\rangle\) & 60 & 140 & 200 \\ 
\midrule 
L1 accept rate (maximum) & \(100\,\mathrm{kHz}\) & \(500\,\mathrm{kHz}\) & \(750\,\mathrm{kHz}\) \\ 
Event Size at HLT input & \(2.0\,\mathrm{MB}\) & \(6.1\,\mathrm{MB}\) & \(8.4\,\mathrm{MB}\) \\ 
Event Network throughput & \(1.6\,\mathrm{Tb} / \mathrm{s}\) & \(24\,\mathrm{Tb} / \mathrm{s}\) & \(51\,\mathrm{Tb} / \mathrm{s}\) \\ 
Event Network buffer (60\,s) & \(12\,\mathrm{TB}\) & \(182\,\mathrm{TB}\) & \(379\,\mathrm{TB}\) \\ 
HLT accept rate & \(1\,\mathrm{kHz}\) & \(5\,\mathrm{kHz}\) & \(7.5\,\mathrm{kHz}\) \\ 
HLT computing power & \(0.7\,\mathrm{MHS} 06\) & \(17\,\mathrm{MHS} 06\) & \(37\,\mathrm{MHS} 06\) \\ 
Event Size at HLT output & \(1.4\,\mathrm{MB}\) & \(4.3\,\mathrm{MB}\) & \(5.9\,\mathrm{MB}\) \\
Storage throughput & \(2\,\mathrm{GB} / \mathrm{s}\) & \(24\,\mathrm{GB} / \mathrm{s}\) & \(51\,\mathrm{GB} / \mathrm{s}\) \\ 
Storage throughput (Heavy-Ion) & \(12\,\mathrm{GB} / \mathrm{s}\) & \(51\,\mathrm{GB} / \mathrm{s}\) & \(51\,\mathrm{GB} / \mathrm{s}\) \\ 
Storage capacity needed (1 day) & \(0.2\,\mathrm{PB}\) & \(1.6\,\mathrm{PB}\) & \(3.3\,\mathrm{PB}\) \\ 
\bottomrule
\end{tabular}}
\caption{\label{tab:1} CMS Phase-2 trigger and DAQ projected running 
parameters, compared to the 
design values of the current (Phase-1) system.
}
\end{table}
\vspace{-1.0em}

\section{Physics Objects and HLT Conceptual Structure}

In the HLT, the reconstruction of the physics objects
-- leptons, hadronic jets and global event quantities --
uses the same algorithms and framework as the offline reconstruction,
the CMS Software Framework (CMSSW){,}
the difference being that the online reconstruction has added emphasis in execution speed.
The reconstruction algorithms are encapsulated in modules, 
which are organised in logical sequences called \emph{HLT paths}.
An HLT path targets a given final state and comprises both 
\emph{producer} modules, which calculate a given quantity of interest,
and
\emph{filter} modules, which interrupt further processing of the event if it does not achieve a certain selection {criteria}.
The HLT paths are then designed around the concept of \emph{early filtering} --
slower, more accurate modules are run only if the event is pre-accepted by faster, less accurate ones.
The \emph{HLT menu} is then the set of all HLT paths used to collect data.
An event is saved for offline analysis if any path accepts it.
It is integrated in such a way that computed variables can be reused amongst paths.
Since CMSSW has had multithreading capabilities since 2016,
the HLT menu allows both parallel event processing and simultaneous execution of modules within a single event.

\subsection{Online Reconstruction}

The greatest difference of Phase-2 {online} reconstruction with respect to the current one is
the reconstruction of energy deposits and charged particle tracks in
the High-Granularity Calorimeter (HGCAL)
and
the Phase-2 Tracker,
which are respectively completely new and redesigned and for the upgraded detector.
These subdetectors have very high granularity,
and their objects reconstruction is usually based {on} iterative procedures:
Kalman filter with deterministic annealing for the tracking  and 
a pair of two algorithms, CLUsters of Energy (CLUE) and The Iterative CLustering (TICL)~\cite{DiPilato:2020mqs} 
for the HGCAL reconstruction.
Those algorithms have to be finely tuned for the constraints imposed by the online environment;
the online Phase-2 baseline tracking is based on only two iterations,
whilst the online HGCAL reconstruction runs four iterations. 

\begin{figure}[htbp]
\begin{center}
\includegraphics[width=0.305\textwidth]{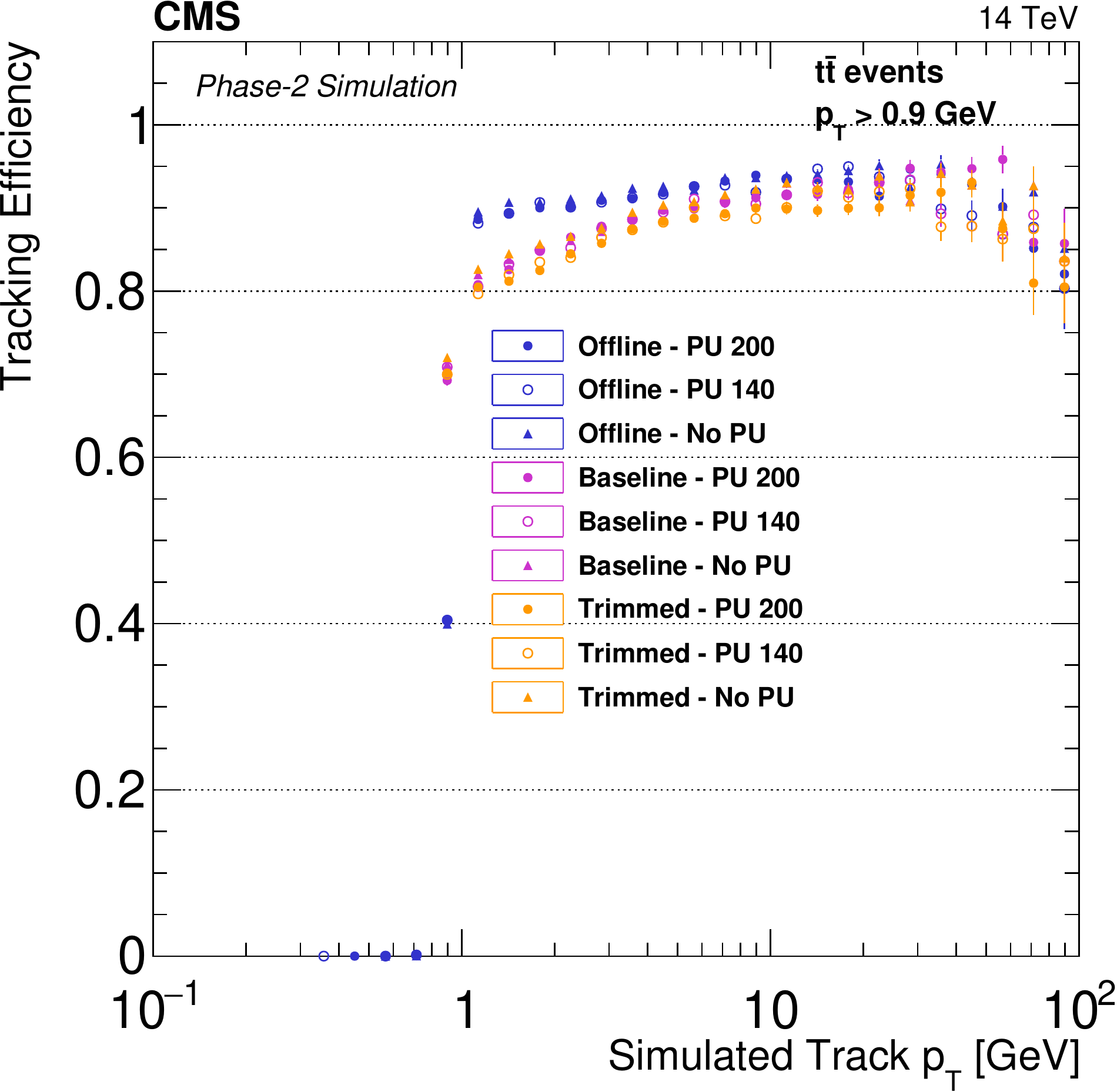}
\includegraphics[width=0.305\textwidth]{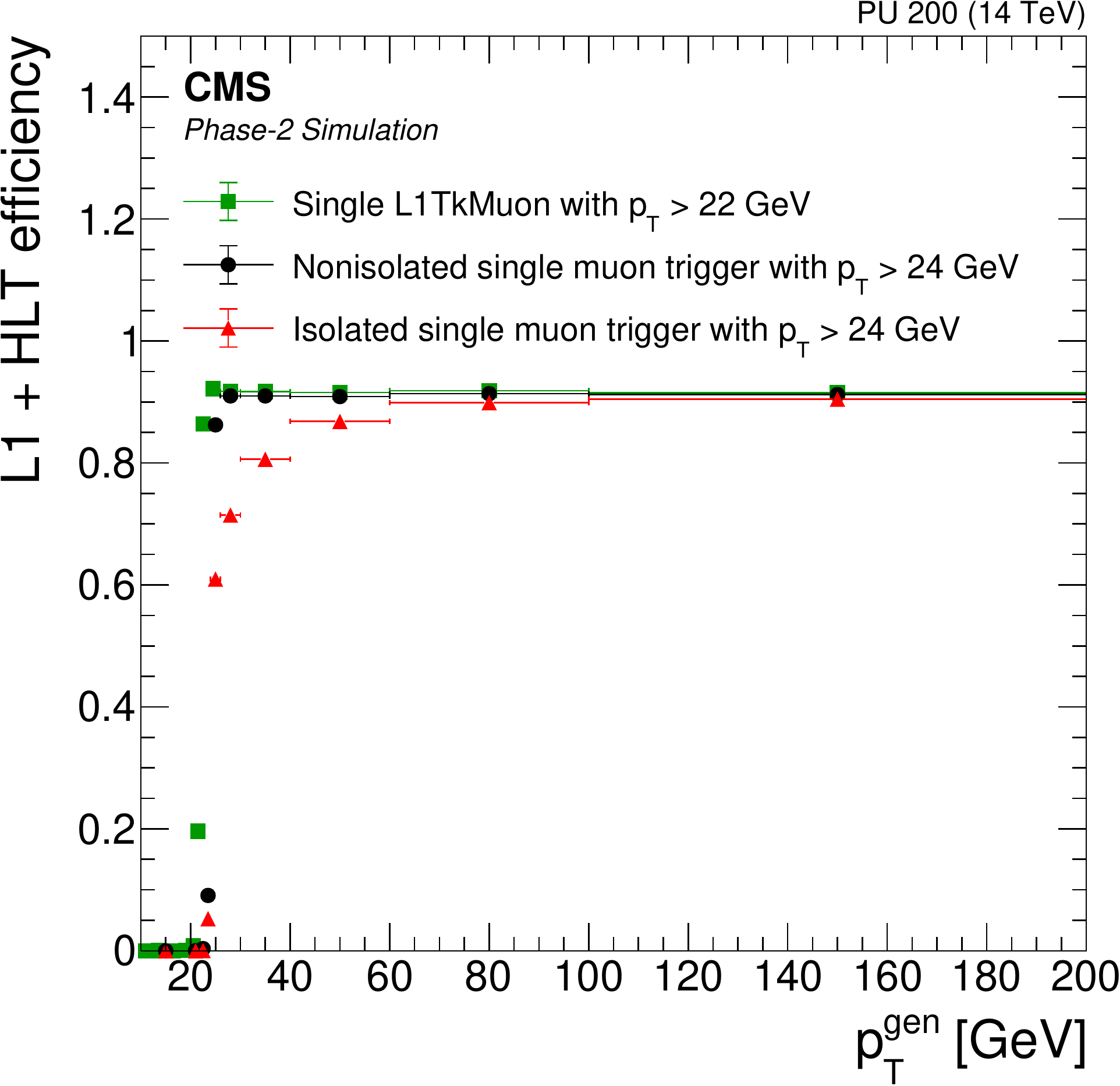}
\includegraphics[width=0.35\textwidth]{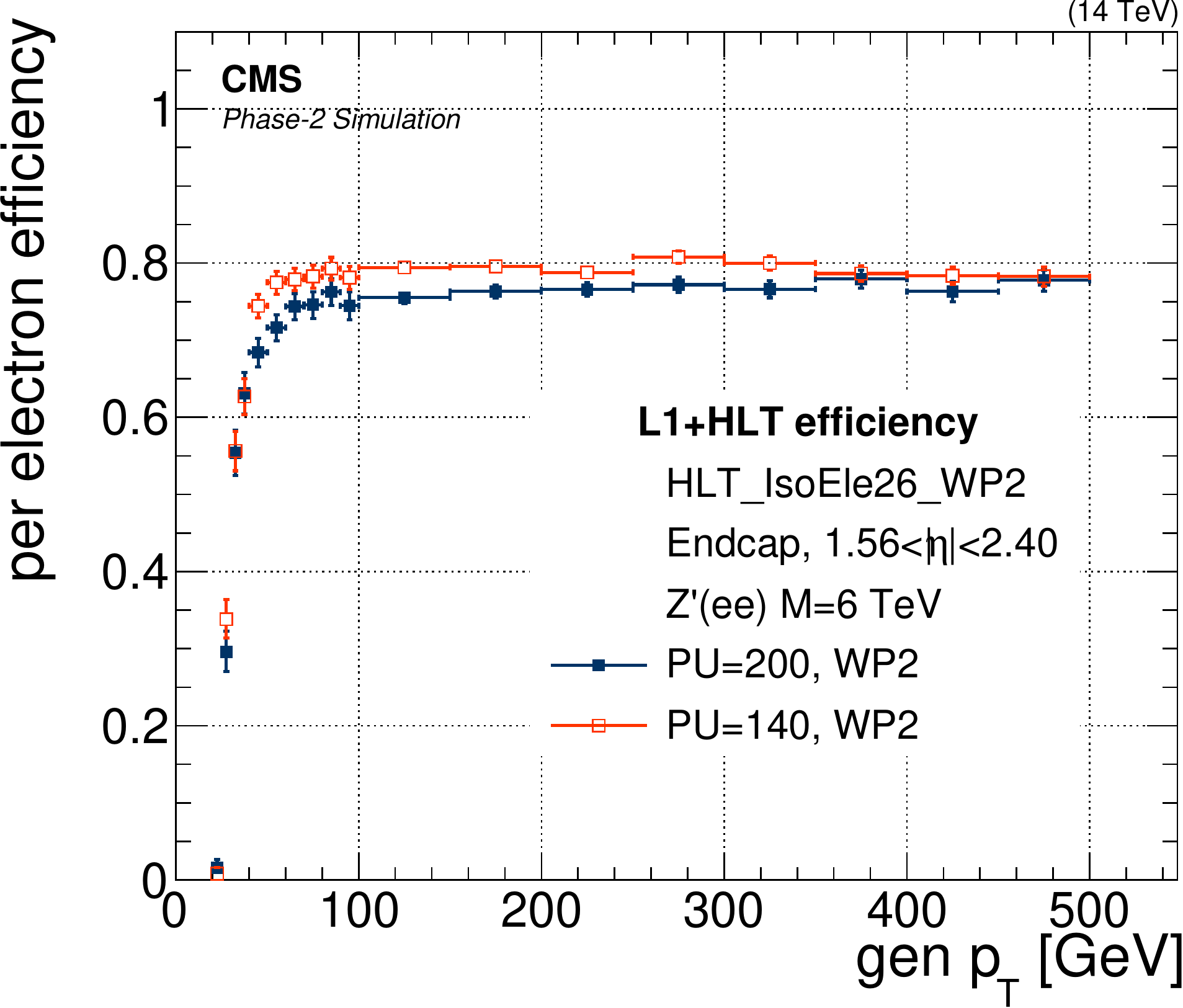}\\
\caption{Efficiency curves for the physics objects: charged particle tracks (left), muons (middle), and electrons (right).
Efficiencies are given as function of the true object parameter, in this case the object \PT. In the case of double-object paths, the efficiency for a single object reconstruction is shown. Efficiencies are measured in adequate signal samples. Full details are available in the DAQ--HLT TDR~\cite{Collaboration:2759072}.}
\label{fig:objects1}
\end{center}
\end{figure}

\begin{figure}[htbp]
\centering
\includegraphics[width=0.35\textwidth]{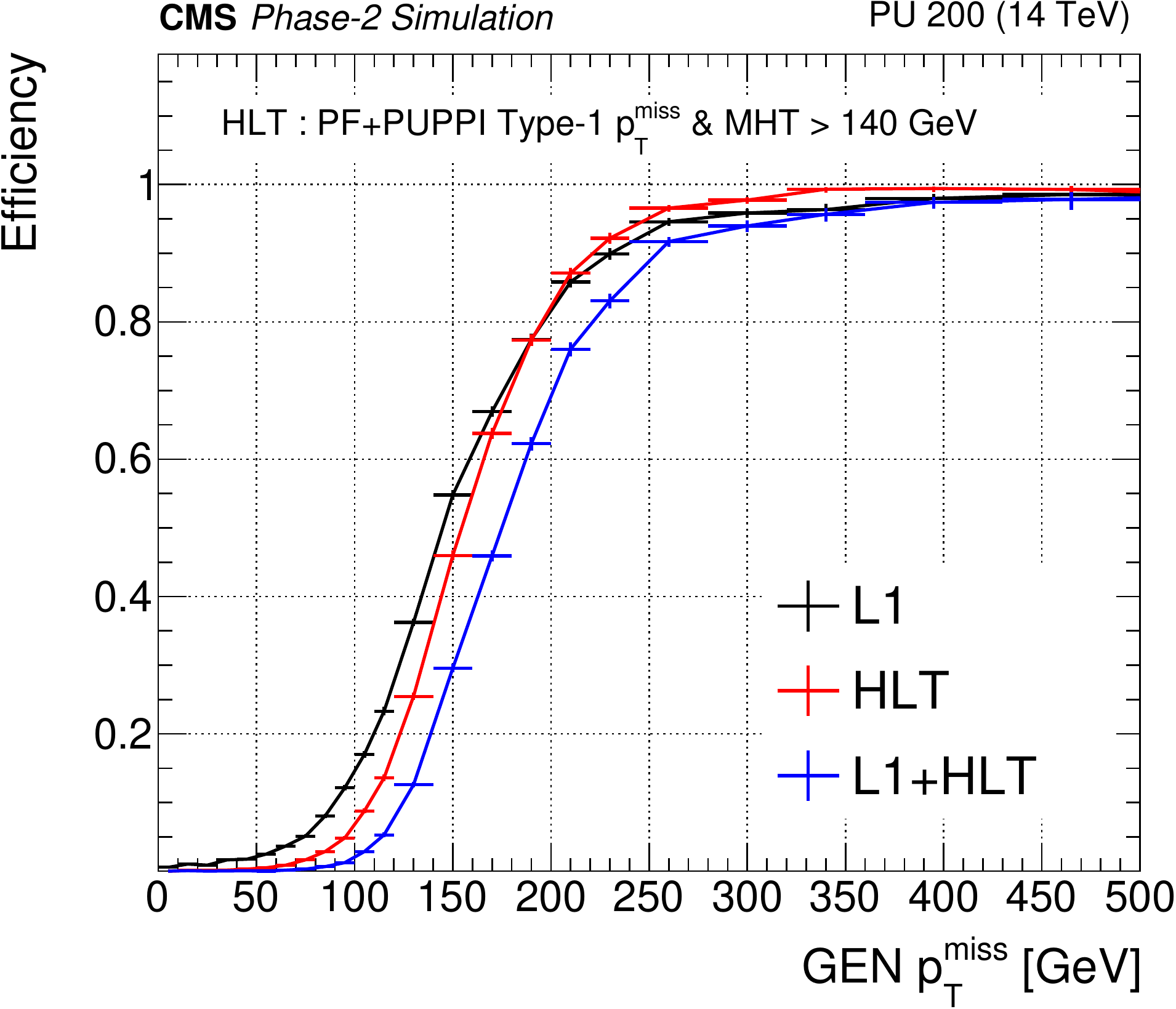}
\includegraphics[width=0.31\textwidth]{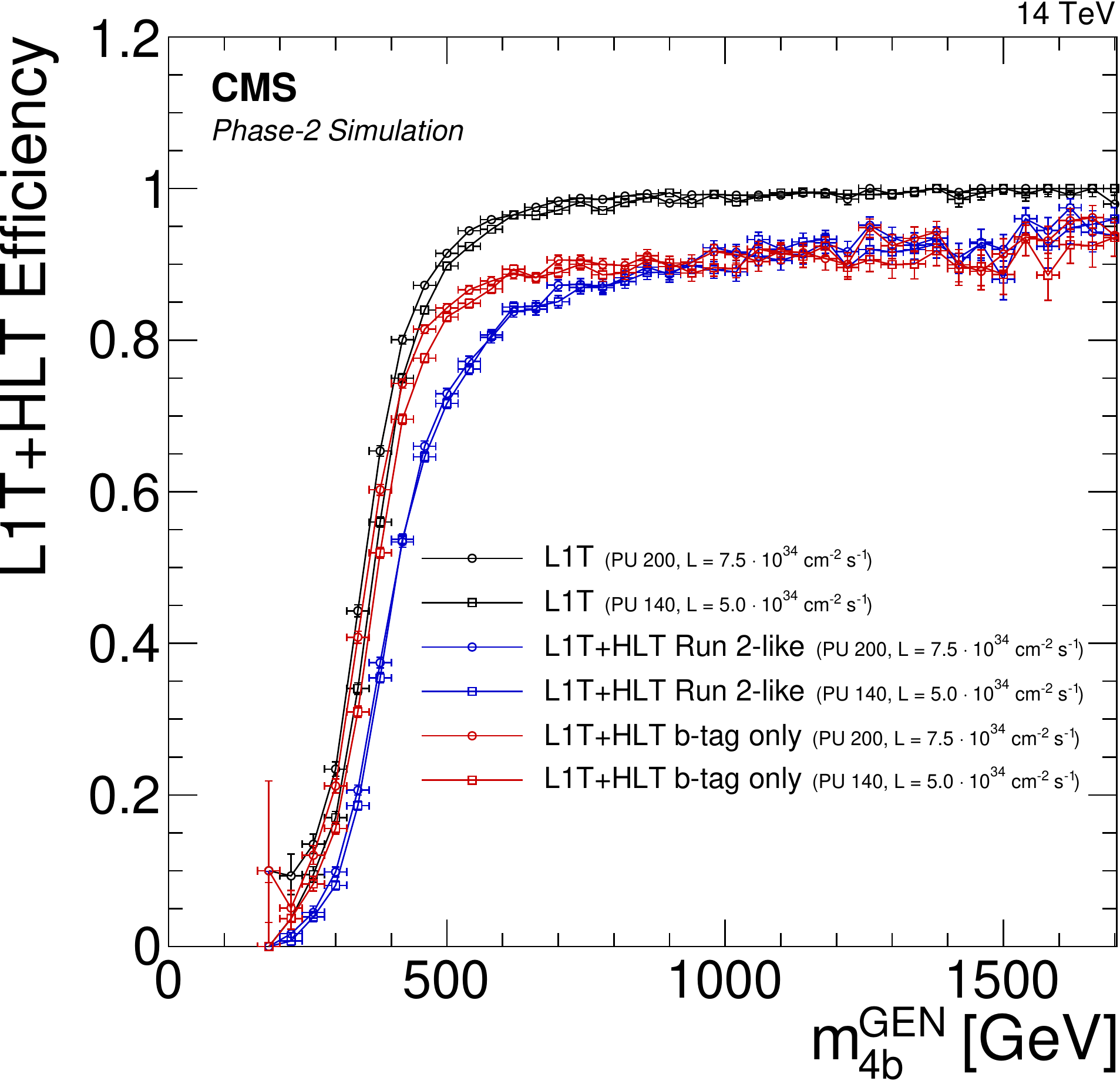}
\includegraphics[width=0.30\textwidth]{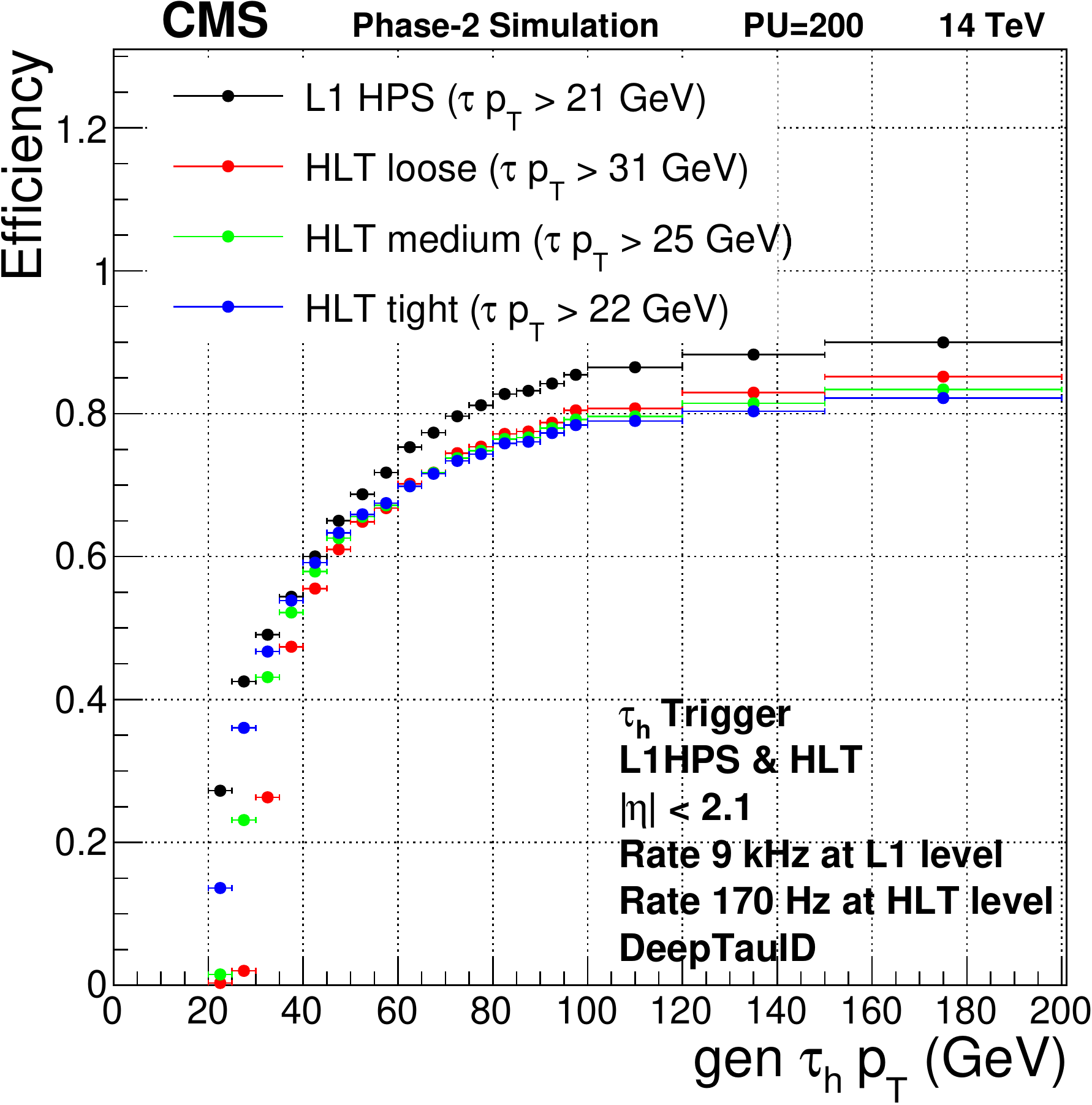}
\caption{Efficiency curves for the physics objects: missing \PT (left), b-tagged jets (middle) and hadronically-decaying tau leptons (right). Efficiencies are given as function of the true object parameter -- usually the object \PT, {except for the b tagging path, where it is the invariant mass of 4 b quarks}. In the case of double-object paths, the efficiency for a single object reconstruction is shown. Efficiencies are measured in adequate signal samples. Full details are available in the DAQ--HLT TDR~\cite{Collaboration:2759072}.}
\label{fig:objects2}
\end{figure}

In the Phase-2 HLT, the reconstruction of the physics objects follows along the same lines as in Phase-1.
All objects are seeded from the results of the L1T reconstruction.
Electrons and photons are seeded by energy deposits in both the electromagnetic calorimeter and in HGCAL;
extensive identification (ID) techniques are employed to reduce backgrounds, 
based on shower shapes and $E/p$ (for electrons only).
Since the Phase-2 L1T has access to Tracker information,
the ``L1TkMuon'' that seeds the HLT muon reconstruction is more precise than the Phase-1 equivalent,
which helps increase the overall reconstruction efficiency.
Hadronic jets and energy sums (missing \PT and total jet transverse momentum \HT) are calculated with inputs from the particle flow algorithm in the barrel and from TICL in the endcap.
Extensive pileup mitigations are needed for these objects, in view of the HL-LHC conditions;
the PUPPI technique~\cite{CMS:2020ebo} was retuned and deployed for the algorithms that reconstruct jets and sums.
Finally, hadronically-decaying tau leptons and b-tagged jets are identified using machine learning techniques,
the DeepTau~\cite{CMS:2022prd} and DeepCSV~\cite{CMS:2017wtu} algorithms.
Figures~\ref{fig:objects1} and \ref{fig:objects2} show efficiency curves for different physics objects in the Phase-2 HLT.

\vspace{-2.5em}
\section{Simplified HLT Menu}
The full Phase-1 HLT menu, as deployed during the 2018 run, comprised approximately 600 HLT paths.
The majority of those paths had very low rate, $\mathcal{O}(1)$ Hz; 
on the other hand, a few paths, like those dedicated to the collection of single, isolated electrons and muons had a much higher rate allocation, $\mathcal{O}(100)$ Hz.
Currently, CMS has prepared a \emph{simplified menu} that targets 50\% of the Phase-2 rate.
It comprises 15 single and double-object based paths, and follows the same overall structure {as} the Phase-1 menu.

The characterisation of an HLT menu involves measuring its execution time and its output rate.
For the rate estimation, we use simulated samples of multijet QCD and electroweak boson events.
Rates are calculated as function of the \PT or ID variable threshold, 
and the overall rate of the menu is calculated summing over all samples.
This procedure double-count{s} events that are selected by more than one {path
but is} adequate for the level of precision intended at this point.
Table~\ref{tab:menu} shows the paths that comprise the simplified menu,
together with their \PT thresholds, their L1T seeding, and their rate percentages (for Phase-1)
and estimated rates (for Phase-2).

%%%%%%
\begin{table}
\hspace{0.10\textwidth}%
\resizebox{0.80\textwidth}{!}{% It cannot overflow even a bit...
\small
 \begin{tabular}{@{} lrrcrrr @{}} % Column formatting, @{} suppresses leading/trailing space
        \toprule
        Trigger type    & \multicolumn{2}{c}{Phase-1} & \hspace{0.5em} & \multicolumn{3}{c}{Phase-2}\\
        \cmidrule{2-3} \cmidrule{4-7}
                        & \shortstack{Threshold\\[3pt][GeV]}
                        & \% rate &
                        & \shortstack{Threshold\\[3pt][GeV]} 
                        & \shortstack{Rate at\\[3pt]\pileupof{140} [Hz]} 
                        & \shortstack{Rate at\\[3pt]\pileupof{200} [Hz]}\\
                \midrule
        Single $\mu$                 & 50            	& 3\% &     &          50            & $155 \pm 6$   &   $213    \pm  8$ \\
        Single $\mu$ (isol.)         & 24            	& 14\% &    &          24            & $943 \pm 32$  &   $1\,111 \pm 29$ \\
        Double $\mu$                 & 37, 27        	& 1\% &     &       37, 27        &  $27 \pm 1$   &   $40     \pm  1$ \\
        Double $\mu$ (isol.)         & 17, 8         	& 2\% &     &       17, 8         & $113 \pm 11$  &   $143    \pm 13$ \\
        Triple $\mu$                 & 5, 3, 3       	& 0.5\% &   &     	10, 5, 5      &  $39 \pm 8$   &   $48     \pm  8$ \\
%%% We keep both electron paths
        Single $e$ (isol.)          & 28            	& 13\% &    &      32 (WP1)      & $609 \pm 27$  &   $1\,005 \pm 33$ \\
                                    &              		&	   &    &      26 (WP2)      & $664 \pm 47$  &   $1\,012 \pm 33$ \\
        Double $e$                  & 25, 25       		& 1\% &     &   25, 25        &  $46 \pm 4$   &   $82     \pm 6$  \\
        Double $e$ (isol.)          & 23, 12        	& 1\% &     &   23, 12        & $52 \pm 5$    &   $104    \pm 9$  \\
        Single $\gamma$                 & 200           & 1\% &     &          187           & $32 \pm 1$    &   $56     \pm 6$  \\
        Single $\gamma$ (isol.)         & 110, EB only  & 1\% &           & 108, EB only  & $35 \pm 9$    &   $52     \pm 7$  \\
        Double $\gamma$                 & 30, 18        & 2\% &       & 30, 23        & $123 \pm 12$  &   $179    \pm 14$ \\
%%%
        Double $\tau$                 & 35, 35        	& 3\% &      & 22, 22        & $106 \pm 18$ &  $159    \pm 27$ \\
%%%
        Single jet                  & 500           	& 1\% &         & 520           & $53 \pm 1$ 	&  $76     \pm  1$ \\
        \HT{}                       & 1050          	& 1\% &          & 1\,070        & $53 \pm 1$ 	&  $74     \pm  1$ \\
        Missing \PT{}               & 120           	& 3\% &          & 140           & $79 \pm 7$    &   $228    \pm 20$ \\
%%% We keep only the Run-2-like path with btag = 0.33
        Multijets                   & \HT = 330     	& 1\% &       & \HT = 330   & $32 \pm 4$     &  $48     \pm 5$  \\
        with b tagging              & jets = 75, 60, 45, 40 	&    &           & jets = 75, 60, 45, 40    &    &              \\
%                                    & 45, 40        	&     &                                & 45, 40        &                &              \\
        \midrule
        \textbf{Total rate}         &               & $\mathbf{49\%}$ & &                   	&	$\mathbf{2\,525 \pm 57}$ & $\mathbf{3\,621 \pm 62}$ \\
        \bottomrule
   \end{tabular}
   }
   \caption{Rates and thresholds of the HLT paths that comprise the simplified menu.
Paths for collection of events with muons, electrons, photons, and multiple b-tagged jets are very similar to those of Phase-1,
whilst those dedicated to regular jets, \HT and missing \PT have only small threshold increases.
The path dedicated to two tau leptons {has had} its thresholds reduced, thanks to the improved capabilities of the L1T.}
   \label{tab:menu}
\end{table}

\begin{figure}[htbp]
\centering
\includegraphics[width=0.4\textwidth]{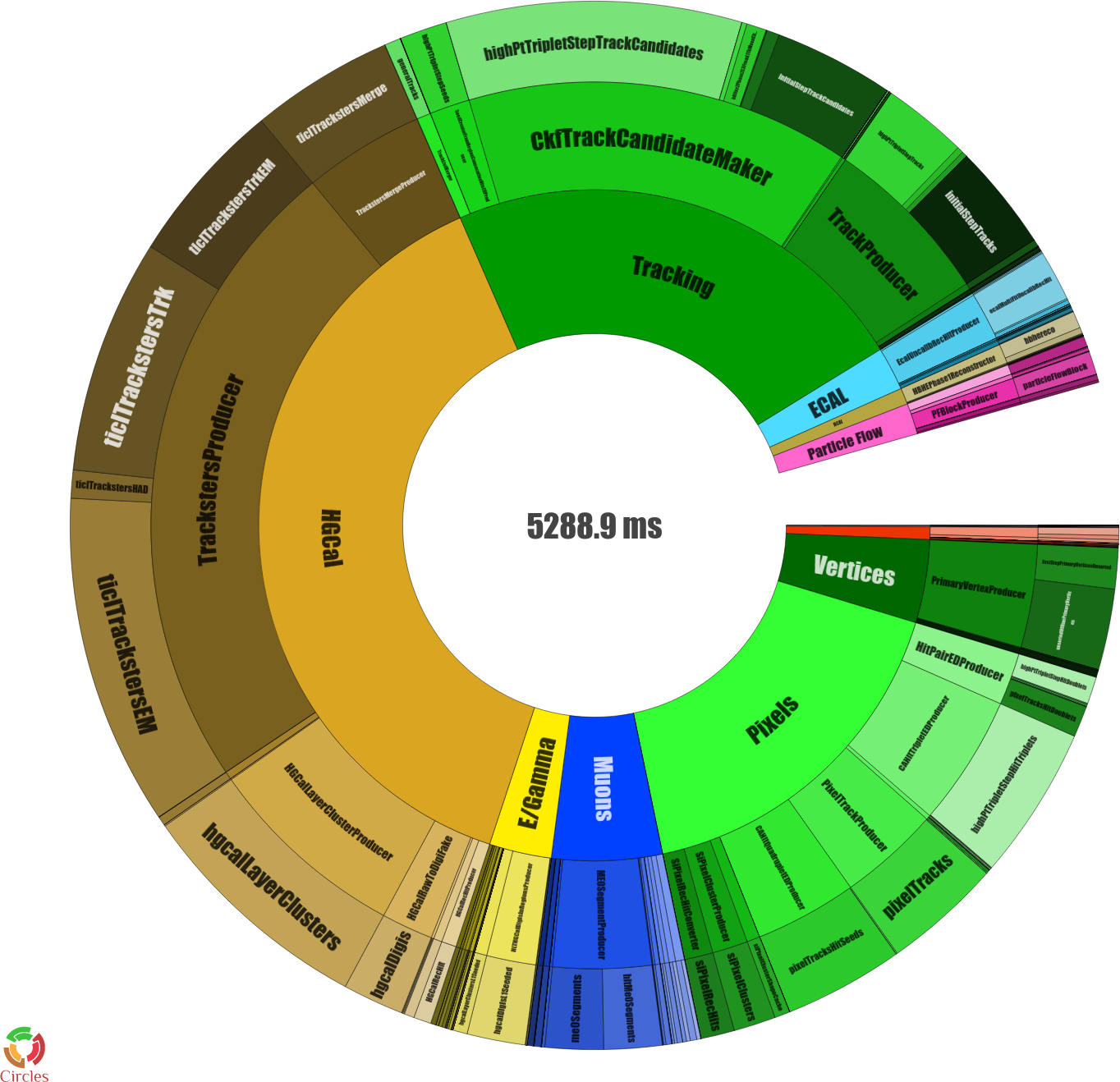}\hspace{4em}
\raisebox{7.6em}{\scalebox{0.8}{\renewcommand{\arraystretch}{0.8}\small\sf\begin{tabular}{lrr}\toprule \textbf{Element} & \textbf{Time} & \textbf{Fraction} \\ \midrule B tagging & 0.4 ms & 0.0 \% \\ E/Gamma & 158.4 ms & 3.0 \% \\ ECAL & 110.9 ms & 2.1 \% \\ Framework & 0.0 ms & 0.0 \% \\ HCAL & 41.6 ms & 0.8 \% \\ HGCal & 2030.5 ms & 38.4 \% \\ HLT & 0.7 ms & 0.0 \% \\ l/O & 0.4 ms & 0.0 \% \\ Jets/MET & 32.1 ms & 0.6 \% \\ L1T & 2.5 ms & 0.0 \% \\ Muons & 280.9 ms & 5.3 \% \\ other & 232.8 ms & 4.4 \% \\ Particle Flow & 78.9 ms & 1.5 \% \\ Pixels & 902.3 ms & 17.1 \% \\ Tracking & 1204.5 ms & 22.8 \% \\ Vertices & 211.9 ms & 4.0 \% \\ \midrule \textit{total} & 5288.9 ms & 100.0 \%\end{tabular}}}
\caption{Measurement of the HLT timing in the reference hardware. 
HGCAL reconstruction (in dark yellow), and pixel and silicon tracking (in green) comprise the majority of the time spent during the online reconstruction.}
\label{fig:timing}
\end{figure}

To estimate the timing of the full menu, 
we extrapolate the measured results of the simplified menu.
We do this by comparing the timing of the equivalent of the simplified menu in 2018 to that of the full one;
after verifying  that the timing distribution is consistent between the two,
we derive {from that comparison} a $+50\%$ correction factor to be applied to the Phase-2 simplified menu timing measurement.
For the measurement itself, we run the HLT menu over a sample of minimum bias samples skimmed with 
a selection consistent with that performed by the Phase-2 L1T.
We use as reference hardware a node with two AMD EPYC 7502 processors, 
totalling 64 (128) physical (logical) cores and a computing power of $\sim1680$\,HS06.
The results are shown in Figure~\ref{fig:timing};
the total timing for the simplified HLT menu is measured to be $\sim5.3$\,s.
Two additional considerations have to be made:
first, we consider improvements to the algorithms {which} could lead, with no performance loss, to a faster HLT by a factor of $1.5\times$--$2\times$.
Second, these results have to be extrapolated to consider the computing hardware that will be available during the Phase-2 era.

\vspace{-0.5em}
\section{Heterogeneous Computing and Extrapolation to Phase-2}
%%% OKAY

We use the term \emph{heterogeneous computing} to refer to the usage of computing nodes equipped with different boards that are better adapted to a given task.
The prototypical example is that of a node equipped with a general-purpose GPU that leverages its high throughput for efficient parallel processing of data.
It must be noted that both the data formats and the algorithm design must be adapted in order to leverage the capability of the heterogenous hardware.

The usage of heterogeneous hardware is a keystone of the CMS Phase-2 computing strategy, both in the offline and online environments.
To start building the expertise needed for the Phase-2 deployment,
CMS has adapted part of its online reconstruction code for heterogeneous computing 
and 
has commissioned an HLT farm with nodes equipped with GPUs starting from 2022.
For the Phase-2 heterogenous HLT, the algorithms for
the HGCAL local reconstruction and
the Patatrack pixel tracking reconstruction~\cite{Bocci:2020pmi}
are already under development.
Assuming that the price/performance ratio of GPUs follows {an evolution similar} to that of CPUs,
and that CMS will be able to adapt 50--80\% of the HLT code to heterogeneous computing,
we estimate that the computing power needs in Table~\ref{tab:1} will be met 
with an effective cost of < 1 CHF/HS06 from 2028 onwards.
%we estimate CMS will be able to deploy an HLT farm with an effective cost of
%0.70 CHF/HS06 in 2028, assuming 50\% of the HLT code is adapted to heterogeneous computing.
%That figure reduces to
%0.22 CHF/HS06 in 2032, with 80\% of the code adapted.

\section{Conclusions}

The CMS Phase-2 online reconstruction is at an advanced stage, and 
a simplified HLT menu based on basic single-object paths {has been} 
developed, integrated and characterised.
The performance of the HLT paths is very close to that of Phase-1,
keeping output rates under control without the need for large increases in
\PT thresholds.
The timing structure of the menu is understood, and
we show it is possible to meet the overall constraints of the Phase-2 system
under two conditions:
improving the speed of the HLT algorithms by a small factor
and
adapting the majority of the code to run on heterogeneous hardware.
We now focus on those two improvements, 
as well as expanding {on} the foundation provided by the simplified menu into the real HLT menu to be deployed in Phase-2.

\acknowledgments
The author would like to thank the São Paulo Research Foundation (FAPESP) for the travel 
funding for this conference, under Grant No. 2018/25225-9.

%\enlargethispage{\baselineskip}

\end{document}